\documentclass[jkps,preprint,fleqn,showpacs,showkeys,amsmath,amssymb]{revtex4}
\usepackage{varioref}
\usepackage{cleveref}
\usepackage{graphicx}
\usepackage{tensor}
\usepackage{bm}
\Crefname{equation}{Eq.}{Eqs.}
\Crefname{section}{Sec.}{Secs.}

\begin{document}

\setcounter{page}{0}
\title[]{A Covariant Approach to 1+3 Formalism}
\author{Chan \surname{Park}}
\email{iamparkchan@kisti.re.kr}
\affiliation{Center for Applied Scientific Computing, Korea Institute of Science and Technology Information, Daejeon, 34141, Korea}

\date[]{Received October 15, 2018}

\begin{abstract}

I present a covariant approach to developing 1+3 formalism without an introduction of any basis or coordinates. In the formalism, a spacetime which has a timelike congruence is assumed. Then, tensors are split into temporal and spatial parts according to the tangent direction to the congruence. I make use of the natural derivatives to define the kinematical quantities and to investigate their properties. They are utilized in the splitting of covariant derivatives. In this way, the Riemann curvature is split into the temporal and spatial part, i.e. Gauss, Codazzi, and Ricci relation. Finally, the splitting of the Einstein equation is achieved by contraction. Choosing congruence as normal to a spacelike hypersurface, the formalism reduces to 3+1 formalism. This approach deepens our understanding of 3+1 formalism. All these processes are performed in a covariant manner without the complexities caused by the introduction of a coordinate system or basis.

\end{abstract}



\maketitle

\section{Introduction}

General relativity is a covariant theory which does not prefer any special coordinates. However, its physical interpretation usually needs a splitting of tensors in time and space. One popular method is splitting by parallel component to timelike vector and orthogonal component to that.\footnote{The orthogonal vector to a timelike direction is spacelike.} Representative examples of applying it include 3+1 formalism and 1+3 formalism \cite{Gourgoulhon2012,GEM,Roy2014}.

In 3+1 formalism, spacetime has a given structure which is a foliation of Cauchy hypersurfaces. Hence, a natural choice of timelike direction is the set of normal vectors to the Cauchy hypersurfaces.\footnote{The normal vectors to a spacelike hypersurface are timelike.} One advantage of it is that the integral curves of normal vectors do not have vorticity by Frobenius' theorem \cite{wald1984general}. Eventually, it greatly simplifies equations of the formalism. Meanwhile, 1+3 formalism assumes a spacetime which has a congruence of timelike curves. Of course, the set of tangent vectors of the curves is a natural choice for the splitting. However, we can not expect to vanish the vorticity of the congruence in general.

Non-vanishing vorticity causes many differences from 3+1 formalism. First of all, the commutator between two spatial vectors is not spatial. It implies that torsion for spatial connection does not vanish. Even it makes the spatial Riemann curvature, in the usual form, not a tensor. Hence, we need another definition of spatial Riemann curvature for spatial connection to make it tensor. In that definition, to be introduced, the spatial Riemann curvature does not satisfy the first Bianchi identity and its Ricci tensor is not symmetric.

The purpose of this paper is deriving the 1+3 splitting of the Einstein equation in a covariant way without introducing any coordinates or bases. There have been, of course, previous studies on 1+3 formalism whose history is well-summarized in\cite{GEM}. Of the notable ones,\cite{GEM,JANTZEN19921} relies on a specific basis and its components in the derivation. Meanwhile,\cite{Roy2014} derives the formalism using general basis, but the introduction of basis complicates the discussion. This paper, also, has differences in the development of the formalism. The kinematical quantities of the congruence are defined by natural derivatives, i.e., Lie derivative and Exterior derivative. Their derived properties will be widely used throughout the paper. On the other hand, the spatial connection is an operation induced from the Levi-Civita connection. Nevertheless, I noticed that it satisfies the condition to be an affine connection and related properties will be investigated.

At first, the splitting of tensors is introduced in \Cref{sec:splitting_of_tensors} \Cref{sec:kinematical_quantities}is devoted to developing the properties of Lie derivative, exterior derivative, and covariant derivative in terms of 1+3 splitting. And kinematical quantities are defined by them. \Cref{sec:spatial_connection}reviews the spatial connection, its torsion, and its Riemann curvature. For linearity of the Riemann curvature, its form is slightly modified as in \cite{GEM,JANTZEN19921,Roy2014}. In \Cref{sec:splitting_of_first_order_derivatives}, I introduce the splitting of covariant derivatives using Lie derivative along the timelike direction and spatial covariant derivatives in a covariant manner. As an application of it, I derive the splitting of the commutator, torsion of the spatial connection, and the spatial Riemann tensor. In \Cref{sec:splitting_of_second_order}, the Riemann curvature is split into temporal and spatial part. Contracting its indices, the splitting of the Einstein equation is yielded in the end. Finally, in \Cref{sec:reducing_to_3+1+formalism}, 3+1 formalism is derived from the results. This paper use abstract indices $a,b,\cdots$ to denote slots of tensors and introduce geometrized unit, $G=1=c$.

\section{Splitting of Tensors}\label{sec:splitting_of_tensors}

Let us consider a spacetime with a smooth timelike vector field $u$ such that
\begin{equation}
    -1=g_{ab}u^{a}u^{b},
\end{equation}
where $g$ is a spacetime metric. One can use the Gram-Schmidt process to split arbitrary vectors $v$ into parallel parts and orthogonal parts to $u$, respectively, which are resulted in
\begin{equation}
    v^{a}=u^{a}\left(-u_{b}v^{b}\right)+\left(\tensor{\delta}{^{a}_{b}}+u^{a}u_{b}\right)v^{b}.
\end{equation}
The first term of right hand side is the parallel part and the second is the orthogonal part. The vector parallel to $u$ is called temporal vector and the one orthogonal to $u$ is called spatial vector. For the sake of brevity, let us introduce the orthogonal projection operator $\bot_{u}$ defined as endomorphism from any vector to spatial vector given by
\begin{align}
    \bot_{u}v^{a}&\equiv\left(\tensor{\delta}{^{a}_{b}}+u^{a}u_{b}\right)v^{b}.
\end{align}
Because it is linear, one can introduce tensor notation with abstract indices as
\begin{equation}
    \tensor{\left[\bot_{u}\right]}{^{a}_{b}}=\tensor{\delta}{^{a}_{b}}+u^{a}u_{b}.
\end{equation}
The operator satisfies that
\begin{align}
    \bot_{u}\left(fu^{a}\right)&=0,\\
    \bot_{u}X^{a}&=X^{a},
\end{align}
where $f$ is a function and $X$ is a spatial vector. From now on, I will denote spatial tensors by capital letter. If an exception occur, I will notify.

Linear forms can also be split based on $u$ in the same way as the vector:
\begin{equation}
    \omega_{a}=u_{a}\left(-u^{b}\omega_{b}\right)+\left(\tensor{\delta}{^{b}_{a}}+u^{b}u_{a}\right)\omega_{b},
\end{equation}
where $\omega$ is a linear form. Hence, one can naturally extend domain of $\bot_{u}$ to linear forms given by
\begin{align}
    \bot_{u}\omega_{a}\equiv\tensor{\left[\bot_{u}\right]}{^{b}_{a}}\omega_{b}.
\end{align}
For tensors with arbitrary rank, $\bot_{u}$ is generalized by
\begin{align}
    \bot_{u}\tensor{t}{^{a_{1}\cdots{}a_{k}}_{b_{1}\cdots{}b_{l}}}=\tensor{t}{^{c_{1}\cdots{}c_{k}}_{d_{1}\cdots{}d_{l}}}\tensor{\left[\bot_{u}\right]}{^{a_{1}}_{c_{1}}}\cdots\tensor{\left[\bot_{u}\right]}{^{a_{k}}_{c_{k}}}\tensor{\left[\bot_{u}\right]}{^{d_{1}}_{b_{1}}}\cdots\tensor{\left[\bot_{u}\right]}{^{d_{l}}_{b_{l}}},
\end{align}
where $t$ is tensor of arbitrary rank. Using the operator, one can express the decomposition of rank $(0,2)$ tensor by
\begin{equation}
    t_{ab}=u_{a}u_{b}\left(u^{c}u^{d}t_{cd}\right)+u_{a}\left(-u^{c}\tensor{\left[\bot_{u}\right]}{^{d}_{b}}t_{cd}\right)+\left(-\tensor{\left[\bot_{u}\right]}{^{c}_{a}}u^{d}t_{cd}\right)u_{b}+\bot_{u}t_{ab}.
\end{equation}

The spatial metric which is a metric for spatial vectors is naturally induced from spacetime metric $g$ by
\begin{align}
    \gamma_{ab}&\equiv\bot_{u}g_{ab}\\
    &=g_{ab}+u_{a}u_{b}.
\end{align}
Actually $\gamma$ is spatial, but I will use a lowercase letter for spatial metric conventionally. Surprisingly,
\begin{align}
    \tensor{\left[\bot_{u}\right]}{^{a}_{b}}=\tensor{\gamma}{^{a}_{b}}.
\end{align}
Despite the spatial metric, I will raise or lower indices by spacetime metric $g$ for all tensor.

\section{Kinematical Quantities}\label{sec:kinematical_quantities}

    Let me investigate properties of Lie derivatives and exterior derivatives which can be defined without covariant derivatives. The kinematical quantities of the congruence are defined with these operations.

    \subsection{Lie Derivatives along $u^{a}$}

    Lie derivative of rank $(0,p)$ spatial tensor along $u^{a}$ is always spatial because
    \begin{align}
        u^{a_{i}}\mathcal{L}_{u}T_{a_{1}\cdots{}a_{i}\cdots{}a_{p}}=\mathcal{L}_{u}\left(u^{a_{i}}T_{a_{1}\cdots{}a_{i}\cdots{}a_{p}}\right)-T_{a_{1}\cdots{}a_{i}\cdots{}a_{p}}\mathcal{L}_{u}u^{a_{i}}=0,
    \end{align}
    for any integer $1\leq{}i\leq{}p$ where $T$ is spatial rank $(0,p)$ tensor. Meanwhile, Lie derivative of a tensor which has contravariant slots is not spatial because $\mathcal{L}_{u}u_{a}$ does not vanish in general. For the sake of brevity, we will denote $\mathcal{L}_{u}T_{a_{1}\cdots{}a_{p}}$ as $\dot{T}_{a_{1}\cdots{}a_{p}}$ from now on:
    \begin{equation}
        \dot{T}_{a_{1}\cdots{}a_{p}}\equiv\mathcal{L}_{u}T_{a_{1}\cdots{}a_{p}}
    \end{equation}
    Note that $\mathcal{L}_{u}T^{a_{1}\cdots{}a_{p}}$ is not equal to $\dot{T}^{a_{1}\cdots{}a_{p}}$, index raising of $\dot{T}$, because $\mathcal{L}_{u}\gamma^{ab}$ does not vanish in general.
    
    Let us define the acceleration $A$ and expansion $\Theta$ as
    \begin{align}
        A_{a}&\equiv\dot{u}_{a},\\
        \Theta_{ab}&\equiv\frac{1}{2}\dot{\gamma}_{ab}.\label{eq:def_expansion}
    \end{align}
    Note that $\Theta$ is spatial because it is Lie derivative of spatial tensor along $u$ and $A$ is also spatial because
    \begin{equation}
        u^{a}A_{a}=-u_{a}\mathcal{L}_{u}u^{a}=0.
    \end{equation}
    Expansion $\Theta$ can be decomposed into trace part and traceless part as
    \begin{equation}
        \Theta_{ab}=\frac{1}{3}\Theta\gamma_{ab}+\Sigma_{ab},
    \end{equation}
    where $\Theta$ is expansion scalar and $\Sigma_{ab}$ is shear.
    Lie derivatives of metric duals of $\gamma$ are yielded by
    \begin{align}
        \mathcal{L}_{u}\gamma^{ab}&=\mathcal{L}_{u}g^{ab}=-\dot{g}^{ab}=u^{a}A^{b}+A^{a}u^{b}-2\Theta^{ab},\label{eq:Lie_derivative_of_gamma_raised}\\
        \mathcal{L}_{u}\tensor{\gamma}{^{a}_{b}}&=\mathcal{L}_{u}\left(g^{ac}\gamma_{cb}\right)=u^{a}A_{b}.
    \end{align}

    \subsection{Exterior Derivative}

    In this section, I introduce index-free notation for brevity denoting differential forms as bold letters. A differential form contracted with $u$ more than once vanishes because the differential form is totally antisymmetric. Hence, the differential form has 1+3 splitting given by
    \begin{equation}
        \bm{s}=\bm{u}\wedge\bm{S}^{\left(\mathrm{E}\right)}+\bm{S}^{\left(\mathrm{M}\right)}
    \end{equation}
    where $\bm{u}$ is the metric dual of $u$, $\bm{s}$ is a $p$-form, $\bm{S}^{\left(\mathrm{E}\right)}$ is the $\left(p-1\right)$-form called by electric part and $\bm{S}^{\left(\mathrm{M}\right)}$ is the $p$-form called by magnetic part. $\bm{S}^{\left(\mathrm{E}\right)}$ and $\bm{S}^{\left(\mathrm{M}\right)}$ are spatial by definition. Each part is yielded by
    \begin{align}
        \bm{S}^{\left(\mathrm{E}\right)}&=-\bm{u}\cdot\bm{s},\\
        \bm{S}^{\left(\mathrm{M}\right)}&=\bot_{u}\bm{s},
    \end{align}
    where binary operator $\cdot$ means contraction between right most slot of left hand side and left most slot of right hand side.
    
    Exterior derivative of $\bm{s}$ is split by
    \begin{equation}
        d\bm{s}=\bm{u}\wedge\left\{-\mathcal{L}_{u}\bm{s}+d\left(\bm{u}\cdot{}\bm{s}\right)\right\}+\bot_{u}d\bm{s}\label{eq:split_of_d_form},
    \end{equation}
    where we use Cartan identity which is
    \begin{equation}
        \mathcal{L}_{u}\bm{s}=\bm{u}\cdot{}d\bm{s}+d\left(\bm{u}\cdot\bm{s}\right).
    \end{equation}
    Thus $d\bm{u}$ is split as
    \begin{equation}
        d\bm{u}=-\bm{u}\wedge\bm{A}-2\bm{\Omega}\label{eq:split_of_u},
    \end{equation}
    where we define vorticity $\bm{\Omega}$ by
    \begin{equation}
        \bm{\Omega}\equiv-\frac{1}{2}\bot_{u}d\bm{u}.\label{eq:def_vorticity}
    \end{equation}
    Meanwhile, exterior derivative has explicit formula as
    \begin{equation}
        d\bm{u}\left(v,w\right)=v\left(\bm{u}\left(w\right)\right)-w\left(\bm{u}\left(v\right)\right)-\bm{u}\left(\left[v,w\right]\right),
    \end{equation}
    where $v$, $w$ are arbitrary vectors. Remind that differential form is a function of vectors and vector is a function of scalar. If we choose spatial vector $X$ and $Y$ to contract $d\bm{u}$ then we get the temporal part of commutator between spatial vectors as
    \begin{equation}
        d\bm{u}\left(X,Y\right)=-\bm{u}\left(\left[X,Y\right]\right)
        =-2\bm{\Omega}\left(X,Y\right),
    \end{equation}
    where second equality comes from \Cref{eq:split_of_u}. If vorticity does not vanish, the commutator between spatial vectors is not spatial in general:
    \begin{equation}
        \bm{u}\left(\left[X,Y\right]\right)
        =2\bm{\Omega}\left(X,Y\right)\label{eq:temporal_part_of_commutator}
    \end{equation}
    This fact is a crucial branch point of 1+3 formalism from 3+1 formalism. I will discuss derived differences throughout the paper.
    
    Twice exterior derivative of $\bm{u}$ vanishes because of Poincar\'e lemma. It can be split as 
    \begin{equation}
        0=d^{2}\bm{u}=\bm{u}\wedge\left(2\mathcal{L}_{u}\bm{\Omega}+\bot_{u}d\bm{A}\right)+2\left(\bm{\Omega}\wedge\bm{A}-\bot_{u}d\bm{\Omega}\right),
    \end{equation}
    where I use direct substitution of \Cref{eq:split_of_u} and splitting of exterior derivative of acceleration and vorticity using \Cref{eq:split_of_d_form}. This is a set of geometrical constraints on the first order derivatives of vorticity given by
    \begin{align}
        \mathcal{L}_{u}\bm{\Omega}&=-\frac{1}{2}\bot_{u}d\bm{A},\label{eq:dot_vorticity}\\
        \bot_{u}d\bm{\Omega}&=\bm{\Omega}\wedge\bm{A}.
    \end{align}
    
    \subsection{Covariant Derivative}
    
    Although Lie derivative and exterior derivative are defined without covariant derivative, there are well-known formulas written in covariant derivative\cite{wald1984general}. Using them, the kinematical quantities are rewritten as
    \begin{align}
        A_{a}&=u^{b}\nabla_{b}u_{a},\\
        \Theta_{ab}&=\frac{1}{2}\bot_{u}\left(\nabla_{b}u_{a}+\nabla_{a}u_{b}\right),\\
        \Omega_{ab}&=\frac{1}{2}\bot_{u}\left(\nabla_{b}u_{a}-\nabla_{a}u_{b}\right).
    \end{align}
    We can see that these are closely related to covariant derivative of $u$. Meanwhile, $\nabla_{b}u_{a}$ has the following property:
    \begin{equation}
        u^{a}\nabla_{b}u_{a}=\frac{1}{2}\nabla_{b}\left(u^{a}u_{a}\right)=0\label{eq:u_nabla_u}
    \end{equation}
    It means that the splitting of $\nabla_{b}u_{a}$ does not have temporal-temporal part and temporal-spatial part. Hence, the covariant derivative is split into
    \begin{equation}
        \nabla_{b}u_{a}=\left(-\tensor{\gamma}{^{c}_{a}}u^{d}\nabla_{d}u_{c}\right)u_{b}+\bot_{u}\nabla_{b}u_{a}=-A_{a}u_{b}+B_{ab},\label{eq:decomposition_of_nabla_u}
    \end{equation}
    where $B$ is defined as
    \begin{equation}
        B_{ab}\equiv\bot_{u}\nabla_{b}u_{a}.
    \end{equation}
    Therefore, we conclude that expansion $\Theta$ and vorticity $\Omega$ are symmetric and antisymmetric part of $B$, respectively, represented by
    \begin{align}
        \Theta_{ab}&=B_{\left(ab\right)},\\
        \Omega_{ab}&=B_{\left[ab\right]},\\
        B_{ab}&=\Theta_{ab}+\Omega_{ab}.
    \end{align}
    It explains why I introduce mysterious factor and sign in \Cref{eq:def_expansion,eq:def_vorticity}.

\section{Spatial Connection}\label{sec:spatial_connection}

Let us define the spatial connection $D$:
\begin{equation}
    D_{X}T=\bot_{u}\nabla_{X}T,
\end{equation}
where $\nabla$ is Levi-Civita connection, $T$ is a spatial tensor of any rank and $X$ is a spatial vector. It is an affine connection in the domain of spatial tensors because
\begin{align}
    D_{X}\left(aY+Z\right)=aD_{X}Y+D_{X}Z,\\
    D_{X}\left(fY\right)=fD_{X}Y+Y^{a}D_{X}f,\\
    D_{aX+Y}Z=aD_{X}Z+D_{Y}Z,\\
    D_{fX}Z=fD_{X}Z,\\
    D_{X}f=X\left(f\right),\\
    D_{X}\left(TU\right)=UD_{X}T+TD_{X}U,
\end{align}
where $X,Y,Z$ are spatial vectors, $T,U$ are spatial tensors of any rank, $a$ is a real number and $f$ is a function\cite{stewart1993advanced}. Based on the linearity of argument $X$ in $D$, we can introduce abstract tensor index for $D$ represented as
\begin{equation}
    D_{X}\tensor{T}{^{a}_{b}}=\left[DT\right]\indices{^{a}_{bc}}X^{c}=X^{c}D_{c}\tensor{T}{^{a}_{b}},
\end{equation}
where $T$ is a rank (1,1) tensor which can be replaced by tensor of any rank. For compatibility of $D_{c}\tensor{T}{^{a}_{b}}$ to all vectors, possibly including temporal component, slot $c$ is regarded as spatial and slots $a$ and $b$ are spatial by definition. The spatial connection $D$ is associated to the spatial metric $\gamma_{ab}$ as shown by
\begin{align}
    D_{X}\gamma_{ab}=\bot_{u}\nabla_{X}\left(g_{ab}+u_{a}u_{b}\right)=0.
\end{align}

The torsion of the spatial connection is given by
\begin{align}
    \tensor[^{\left(3\right)}]{T}{}\left(X,Y\right)&=D_{X}Y-D_{Y}X-\left[X,Y\right]\label{eq:torsion},
\end{align}
where $X$, $Y$ are spatial vectors. It is linear for both arguments $X,Y$ which can be shown easily. Hence, \Cref{eq:torsion} is tensor. And I introduce abstract index notation for the torsion such that
\begin{equation}
    \left[\tensor[^{\left(3\right)}]{T}{}\left(X,Y\right)\right]^{a}=\tensor[^{\left(3\right)}]{T}{^{a}_{bc}}X^{b}Y^{c}.
\end{equation}
For compatibility of $\tensor[^{\left(3\right)}]{T}{^{a}_{bc}}$ to all vectors or linear forms, possibly including temporal component, slot $b$ and $c$ is regarded as spatial such that its action to arbitrary vector becomes action of orthogonal projection of vector:
\begin{equation}
    \tensor[^{\left(3\right)}]{T}{^{c}_{ab}}v^{a}w^{b}=\tensor[^{\left(3\right)}]{T}{^{c}_{ab}}\tensor{\gamma}{^{a}_{d}}\tensor{\gamma}{^{b}_{e}}v^{d}w^{e}
\end{equation}
where $v$ and $w$ are arbitrary vectors. Slots $a$, however, can be any vector because $[X,Y]^{a}$ is not spatial in general even though $X,Y$ are spatial as shown by \Cref{eq:temporal_part_of_commutator}. Eventually, the torsion does not vanish unlike the torsion of Levi-Civita connection. From the expansion of \Cref{eq:torsion}, we get
\begin{equation}
    \tensor[^{\left(3\right)}]{T}{^{c}_{ab}}\left(df\right)_{c}=D_{b}D_{a}f-D_{a}D_{b}f,\label{eq:torsion_expansion}
\end{equation}
where $f$ is a function.

One can define naively the Riemann curvature of the spatial connection as follows:
\begin{equation}
    \tensor[^{\left(3\right)}]{\bar{R}}{}\left(X,Y\right)Z\equiv{}D_{X}D_{Y}Z-D_{Y}D_{X}Z-D_{\left[X,Y\right]}Z\label{eq:naive_Riemann_curvature}
\end{equation}
where $X,Y,Z$ are spatial vectors. However, it is not tensor due to the non-linearity for argument $Z$ as you can see that
\begin{equation}
    \tensor[^{\left(3\right)}]{\bar{R}}{}\left(X,Y\right)\left(fZ\right)=f\bar{R}\left(X,Y\right)Z+Z\left(\left[X,Y\right]\left(f\right)-D_{\left[X,Y\right]}f\right).
\end{equation}
Note that the second term of the right hand side does not vanish because $\left[X,Y\right]$ has a temporal component in general shown in \Cref{eq:temporal_part_of_commutator}. Alternatively, we can consider the following as a Riemann curvature of the spatial connection:
\begin{equation}
    \tensor[^{\left(3\right)}]{R}{}\left(X,Y\right)Z\equiv{}D_{X}D_{Y}Z-D_{Y}D_{X}Z-\bot_{u}\nabla_{\left[X,Y\right]}Z\label{eq:3d_riemann_curvature}
\end{equation}
Indeed, it is linear for all arguments $X,Y,Z$ which can be shown easily. Hence, \Cref{eq:3d_riemann_curvature} is tensor. And I introduce abstract index notation for the Riemann curvature tensor such that
\begin{equation}
    \left[\tensor[^{\left(3\right)}]{R}{}\left(X,Y\right)Z\right]^{a}=\tensor[^{\left(3\right)}]{R}{^{a}_{bcd}}Z^{b}X^{c}Y^{d}.
\end{equation}
For compatibility of  $\tensor[^{\left(3\right)}]{R}{^{a}_{bcd}}$ to all vectors, possibly including temporal component, slots $b,c,d$ are regarded as spatial and slot $a$ is spatial by definition. From the expansion of \Cref{eq:3d_riemann_curvature}, we get
\begin{equation}
    \tensor[^{\left(3\right)}]{R}{^{a}_{bcd}}Z^{b}=D_{c}D_{d}Z^{a}-D_{d}D_{c}Z^{a}+\tensor{\gamma}{^{a}_{e}}\tensor[^{\left(3\right)}]{T}{^{b}_{cd}}\nabla_{b}Z^{e}.
\end{equation}
The definition of \Cref{eq:3d_riemann_curvature} first appeared in \cite{Massa1974_1,Massa1974_2,Massa1974_3} and was studied in \cite{GEM,JANTZEN19921,boersma1995,Roy2014}.

\section{Splitting of Covariant Derivatives}\label{sec:splitting_of_first_order_derivatives}

We will split first and second order covariant derivatives into the temporal part and spatial part. The derivation will be shown in detail to introduce the algebra.

    \subsection{First Order Covariant Derivatives}
    
    To split covariant derivatives along $u^{a}$, one can use Lie derivatives that are resulted in
    \begin{equation}
        u^{b}\nabla_{b}T_{a_{1}\cdots{}a_{p}}=\mathcal{L}_{u}T_{a_{1}\cdots{}a_{p}}-\sum_{i=1}^{p}T_{a_{1}\cdots{}c\cdots{}a_{p}}\nabla_{a_{i}}u^{c}=\dot{T}_{a_{1}\cdots{}a_{p}}+\sum_{i=1}^{p}T_{a_{1}\cdots{}c\cdots{}a_{p}}\left(u_{a_{i}}A^{c}-\tensor{B}{^{c}_{a_{i}}}\right).\label{eq:decomposition_of_covariant_derivative}
    \end{equation}
    where $T$ is rank $(0,p)$ spatial tensor and I use \Cref{eq:decomposition_of_nabla_u} in the derivation. The splitting of covariant derivatives along arbitrary direction is achieved from the spatial covariant derivatives:
    \begin{align}
        D_{b}T_{a_{1}\cdots{}a_{p}}&=\tensor{\gamma}{^{c_{1}}_{a_{1}}}\cdots\tensor{\gamma}{^{c_{p}}_{a_{p}}}\left(\tensor{\delta}{^{d}_{b}}+u^{d}u_{b}\right)\nabla_{d}T_{c_{1}\cdots{}c_{p}}\\
        &=\tensor{\gamma}{^{c_{1}}_{a_{1}}}\cdots\tensor{\gamma}{^{c_{p}}_{a_{p}}}\nabla_{b}T_{c_{1}\cdots{}c_{p}}+u_{b}\left(-\sum_{i=1}^{p}T_{a_{1}\cdots{}c\cdots{}a_{p}}\tensor{B}{^{c}_{a_{i}}}+\dot{T}_{a_{1}\cdots{}a_{p}}\right)\\
        &=\tensor{\gamma}{^{c_{2}}_{a_{2}}}\cdots\tensor{\gamma}{^{c_{p}}_{a_{p}}}\left\{\nabla_{b}T_{a_{1}c_{2}\cdots{}c_{p}}-T_{c_{1}\cdots{}c_{p}}u_{a_{1}}\left(-u_{b}A^{c_{1}}+\tensor{B}{^{c_{1}}_{b}}\right)\right\}+\cdots\\
        &=\nabla_{b}T_{a_{1}\cdots{}a_{p}}+\sum_{i=1}^{p}T_{a_{1}\cdots{}c\cdots{}a_{p}}\left(u_{a_{i}}u_{b}A^{c}-u_{a_{i}}\tensor{B}{^{c}_{b}}-u_{b}\tensor{B}{^{c}_{a_{i}}}\right)+u_{b}\dot{T}_{a_{1}\cdots{}a_{p}}.
    \end{align}
    Therefore,
    \begin{equation}
        \nabla_{b}T_{a_{1}\cdots{}a_{p}}=\sum_{i=1}^{p}T_{a_{1}\cdots{}c\cdots{}a_{p}}\left(-u_{a_{i}}u_{b}A^{c}+u_{a_{i}}\tensor{B}{^{c}_{b}}+u_{b}\tensor{B}{^{c}_{a_{i}}}\right)-u_{b}\dot{T}_{a_{1}\cdots{}a_{p}}+D_{b}T_{a_{1}\cdots{}a_{p}}.
    \end{equation}
    Let me write some special cases for later:
    \begin{align}
        \nabla_{b}X^{a}&=X^{c}\left(-u^{a}u_{b}A_{c}+u^{a}B_{cb}+u_{b}\tensor{B}{_{c}^{a}}\right)-u_{b}\dot{X}^{a}+D_{b}X^{a}\label{eq:deomposition_of_nabla_spatial_vector}\\
        \nabla_{a}X^{a}&=X^{a}A_{a}+D_{a}X^{a}\label{eq:decomposition_of_div_vector}\\
        \nabla_{c}Y_{ab}&=Y_{db}\left(-u_{a}u_{c}A^{d}+u_{a}\tensor{B}{^{d}_{c}}+u_{c}\tensor{B}{^{d}_{a}}\right)+Y_{ad}\left(-u_{b}u_{c}A^{d}+u_{d}\tensor{B}{^{d}_{c}}+u_{c}\tensor{B}{^{d}_{b}}\right)\nonumber\\
        &\quad-u_{c}\dot{Y}_{ab}+D_{c}Y_{ab}\label{eq:decomposition_of_nabla_spatial_tensor}\\
        \nabla^{b}Y_{ab}&=u_{a}Y_{bc}B^{bc}+Y_{ab}A^{b}+D^{b}Y_{ab}\label{eq:decomposition_of_div_tensor}
    \end{align}
    where $X^{a}$ and $Y_{ab}$ are spatial tensors.
    Notice that $\dot{X}^{a}$ is index raising of $\mathcal{L}_{u}X_{a}$, not $\mathcal{L}_{u}X^{a}$.

    Using the above result, we can get the splitting of commutator between spatial vectors given by
    \begin{align}
        \left[X,Y\right]^{a}&=X^{b}\nabla_{b}Y^{a}-Y^{b}\nabla_{b}X^{a}\\
        &=u^{a}X^{b}Y^{c}B_{cb}+X^{b}D_{b}Y^{a}-u^{a}Y^{b}X^{c}B_{cb}-Y^{b}D_{b}X^{a}\\
        &=D_{X}Y^{a}-D_{Y}X^{a}-2u^{a}\Omega_{bc}X^{b}Y^{c}
    \end{align}
    where $X^{a},Y^{a}$ are spatial vectors. Eventually, we have the torsion and the Riemann curvature associated with the spatial connection yielded as
    \begin{align}
        \tensor[^{\left(3\right)}]{T}{^{a}_{bc}}&=2u^{a}\Omega_{bc},\\
        \tensor[^{\left(3\right)}]{R}{^{a}_{bcd}}Z^{b}&=D_{c}D_{d}Z^{a}-D_{d}D_{c}Z^{a}+2\Omega_{cd}\left(\dot{Z}^{a}-Z^{b}\tensor{B}{_{b}^{a}}\right).
    \end{align}

    \subsection{Second Order Covariant Derivatives}

    Using \Cref{eq:deomposition_of_nabla_spatial_vector,eq:decomposition_of_nabla_u,eq:decomposition_of_nabla_spatial_tensor}, we get some parts of second order covariant derviatives of $u$ and spatial vector $X$ as follows:
    \begin{align}
        \bot_{u}\nabla_{c}\nabla_{b}u_{a}&=D_{c}B_{ab}-A_{a}B_{bc},\label{eq:two_nabla_1}\\
        \bot_{u}\left(u^{b}\nabla_{c}\nabla_{b}u_{a}\right)&=D_{c}A_{a}-B_{ab}\tensor{B}{^{b}_{c}},\label{eq:two_nabla_2}\\
        \bot_{u}\left(u^{c}\nabla_{c}\nabla_{b}u_{a}\right)&=\dot{B}_{ab}-A_{a}A_{b}-B_{cb}\tensor{B}{^{c}_{a}}-B_{ac}\tensor{B}{^{c}_{b}},\label{eq:two_nabla_3}\\
        \bot_{u}\nabla_{c}\nabla_{d}X^{a}&=X^{b}\left(\tensor{B}{^{a}_{c}}B_{bd}+B_{dc}\tensor{B}{_{b}^{a}}\right)-B_{dc}\dot{X}^{a}+D_{c}D_{d}X^{a}.\label{eq:two_nabla_4}
    \end{align}
    Although there are total 8 parts for the splitting of a second order derivative, I represent only necessary parts to the next section.

\section{Einstein Equation}\label{sec:splitting_of_second_order}

Now we are all ready to split the Einstein equation. We first split the Riemann curvature, show the splitting of the Ricci and the Einstein tensor, introduce the splitting of the stress-energy tensor, and split the Einstein equation.

    \subsection{Riemann Curvature}
    
    The Riemann curvature tensor associated with Levi-Civita connection is defined by
    \begin{equation}
        \tensor{R}{}\left(u,v\right)w=\nabla_{u}\nabla_{v}w-\nabla_{v}\nabla_{u}w-\nabla_{\left[u,v\right]}w
    \end{equation}
    where $u,v,w$ are arbitrary vectors. Also, the Riemann curvature tensor $R_{abcd}$, lowering the first index, has symmetries:
    \begin{align}
        R_{\left(ab\right)cd}&=0,&
        R_{ab\left(cd\right)}&=0,&
        R_{a\left[bcd\right]}&=0.\label{eq:symmetry_of_Riemann_curvature}
    \end{align}
    Hence, of the split parts of the tensor, there are only 3 classes of non-vanishing parts which are
    \begin{align}
        \bot_{u}\tensor{R}{^{a}_{bcd}},&&\bot_{u}\left(\tensor{R}{^{a}_{bcd}}u^{b}\right),&&\bot_{u}\left(\tensor{R}{^{a}_{bcd}}u^{b}u^{d}\right).
    \end{align}
    
    Finally, using \Cref{eq:two_nabla_1,eq:two_nabla_2,eq:two_nabla_3,eq:two_nabla_4}, we get
    \begin{align}
        \left(\bot_{u}\tensor{R}{^{a}_{bcd}}\right)X^{b}&=\bot_{u}\left(\nabla_{c}\nabla_{d}-\nabla_{d}\nabla_{c}\right)X^{a}=\left(\tensor[^{\left(3\right)}]{R}{^{a}_{bcd}}+\tensor{B}{^{a}_{c}}B_{bd}-\tensor{B}{^{a}_{d}}B_{bc}\right)X^{b},\label{eq:Gauss_relation}\\
        \bot_{u}\left(\tensor{R}{^{a}_{bcd}}u^{b}\right)&=\bot_{u}\left(\nabla_{c}\nabla_{d}-\nabla_{d}\nabla_{c}\right)u^{a}=2A^{a}\Omega_{cd}+D_{c}\tensor{B}{^{a}_{d}}-D_{d}\tensor{B}{^{a}_{c}},\label{eq:Codazzi_relation}\\
        \bot_{u}\left(\tensor{R}{^{a}_{bcd}}u^{b}u^{d}\right)&=\bot_{u}\left\{u^{d}\left(\nabla_{c}\nabla_{d}-\nabla_{d}\nabla_{c}\right)u^{a}\right\}=D_{c}A^{a}+A^{a}A_{c}+\tensor{B}{^{b}_{c}}\tensor{B}{_{b}^{a}}-\tensor{\dot{B}}{^{a}_{c}}.\label{eq:Ricci_relation}
    \end{align}
    They are callled Gauss, Codazzi, and Ricci relation, respectively. From symmetries of the Riemann curvature tensor associated with the Levi-Civita connection, \Cref{eq:symmetry_of_Riemann_curvature}, and the Gauss relation, \Cref{eq:Gauss_relation}, we get symmetries of $\tensor[^{\left(3\right)}]{R}{_{abcd}}$ as
    \begin{align}
        \tensor[^{\left(3\right)}]{R}{_{\left(ab\right)cd}}&=0,\\
        \tensor[^{\left(3\right)}]{R}{_{ab\left(cd\right)}}&=0,\\
        \tensor[^{\left(3\right)}]{R}{_{a\left[bcd\right]}}&=2B_{a\left[b\right.}\Omega_{\left.cd\right]}.\label{eq:antisymmetrized_Riemann_curvature_of_spatial_connection}
    \end{align}
    Note that, unlike the Riemann curvature associated with the Levi-Civita connection, \Cref{eq:antisymmetrized_Riemann_curvature_of_spatial_connection} does not vanish due to the vorticity.

    Let us define the spatial Ricci tensor as
    \begin{equation}
        \tensor[^{\left(3\right)}]{R}{_{ab}}=\tensor{\gamma}{^{c}_{d}}\tensor[^{\left(3\right)}]{R}{^{d}_{acb}}.
    \end{equation}
    Contracting indices $a$ and $c$ in \Cref{eq:antisymmetrized_Riemann_curvature_of_spatial_connection}, we get the antisymmetric part of the spatial Ricci tensor given by
    \begin{equation}
        \tensor[^{\left(3\right)}]{R}{_{\left[ab\right]}}=-2\tensor{\Omega}{^{c}_{\left[a\right.}}\Theta_{\left.b\right]c}-\Theta\Omega_{ab}.\label{eq:antisymmetric_part_of_spatial_Ricci_tensor}
    \end{equation}
    Notice that $\tensor[^{\left(3\right)}]{R}{_{ab}}$ is not symmetric in general2 because of vorticity.

    \subsection{Einstein Tensor}    
    
    Contracting indices $a$ and $c$ in \Cref{eq:Gauss_relation,eq:Codazzi_relation,eq:Ricci_relation}, we yield splitting of the Ricci tensor resulted in
    \begin{align}
        R_{ab}u^{a}u^{b}&=-\dot{\Theta}-\Theta^{ab}\Theta_{ab}+\Omega^{ab}\Omega_{ab}+D_{a}A^{a}+A^{a}A_{a}\label{eq:temporal_temporal_Ricci}\\
        R_{cb}\tensor{\gamma}{^{c}_{a}}u^{b}&=D_{b}\tensor{B}{^{b}_{a}}-D_{a}\Theta+2A^{b}\Omega_{ba}\label{eq:spatial_temporal_Ricci}\\
        \bot_{u}R_{ab}&=\tensor[^{\left(3\right)}]{R}{_{\left(ab\right)}}+\dot{\Theta}_{ab}-2\tensor{\Theta}{_{a}^{c}}\Theta_{bc}-2\tensor{\Omega}{^{c}_{\left(a\right.}}\Theta_{\left.b\right)c}+\Theta\Theta_{ab}-D_{\left(a\right.}A_{\left.b\right)}-A_{a}A_{b}.\label{eq:spatial_spatial_Ricci}
    \end{align}
    In the derivation of \Cref{eq:temporal_temporal_Ricci}, \Cref{eq:Lie_derivative_of_gamma_raised} was used in contraction of $\dot{B}$. In \Cref{eq:spatial_spatial_Ricci}, \Cref{eq:dot_vorticity,eq:antisymmetric_part_of_spatial_Ricci_tensor} were used to eliminate $\dot{\Omega}$ and the spatial Ricci tensor, respectively. Using \Cref{eq:temporal_temporal_Ricci} and the trace of \Cref{eq:spatial_spatial_Ricci}, we get the Ricci scalar which is
    \begin{equation}
        R=\tensor[^{\left(3\right)}]{R}{}+2\dot{\Theta}+\Theta^{ab}\Theta_{ab}-\Omega^{ab}\Omega_{ab}+\Theta^{2}-2D_{a}A^{a}-2A^{a}A_{a}.
    \end{equation}
    Then, the Einstein tensor is split as
    \begin{align}
        G_{ab}u^{a}u^{b}&=\frac{1}{2}\left(\tensor[^{\left(3\right)}]{R}{}-\Theta^{ab}\Theta_{ab}+\Omega^{ab}\Omega_{ab}+\Theta^{2}\right),\\
        G_{cb}\tensor{\gamma}{^{c}_{a}}u^{b}&=D_{b}\tensor{B}{^{b}_{a}}-D_{a}\Theta+2A^{b}\Omega_{ba},\\
        \bot_{u}G_{ab}&=\frac{1}{3}\bot_{ab}\left(-\frac{1}{2}\tensor[^{\left(3\right)}]{R}{}-2\dot{\Theta}-\frac{3}{2}\Theta_{cd}\Theta^{cd}+\frac{3}{2}\Omega_{cd}\Omega^{cd}-\frac{1}{2}\Theta^{2}+2D_{c}A^{c}+2A^{c}A_{c}\right)\nonumber\\
        &\quad+\tensor[^{\left(3\right)}]{R}{_{\left<ab\right>}}+\dot\Sigma_{\left<ab\right>}-2\tensor{\Sigma}{^{c}_{\left<a\right.}}\Sigma_{\left.b\right>c}-2\tensor{\Omega}{^{c}_{\left(a\right.}}\Sigma_{\left.b\right)c}+\frac{1}{3}\Theta\Sigma_{ab}-D_{\left<a\right.}A_{\left.b\right>}-A_{\left<a\right.}A_{\left.b\right>},
    \end{align}
    where $\left<ab\right>$ denote the traceless part of rank $(0,2)$ symmetrized tensor defined by
    \begin{equation}
        T_{\left<ab\right>}=T_{\left(ab\right)}-\frac{1}{3}\gamma_{ab}\left(\gamma^{cd}T_{cd}\right),
    \end{equation}
    for arbitrary rank $(0,2)$ tensor $T$.

    \subsection{Stress-energy}

    Stress-energy tensor $T_{ab}$ has splitting given by
    \begin{equation}
        T_{ab}=Eu_{a}u_{b}+u_{a}P_{b}+P_{a}u_{b}+S_{ab}, 
    \end{equation}
    where $E$ is the energy density, $P_{a}$ is the momentum density and $S_{ab}$ is the stress with respect to $u^{a}$, respectively. Note that $P_{a}$ and $S_{ab}$ are spatial tensors. The local conservation of $T_{ab}$ has splitting given by
    \begin{align}
        0=\nabla^{b}T_{ab}&=u_{a}\left(\dot{E}+D^{b}P_{b}+E\Theta+2P_{b}A^{b}+S_{bc}\Theta^{bc}\right)\nonumber\\
        &\quad+\dot{P}_{a}+D^{b}S_{ab}+EA_{a}+2P_{b}\tensor{\Omega}{_{a}^{b}}+P_{a}\Theta+S_{ab}A^{b},
    \end{align}
    where I used \Cref{eq:decomposition_of_div_vector,eq:decomposition_of_div_tensor} in the derivation.
    
    \subsection{Einstein Equation}
    
    Finally, the splitting of the Einstein equation $G_{ab}=8\pi{}T_{ab}$ is given by
    \begin{align}
        16\pi{}E&=\tensor[^{\left(3\right)}]{R}{}-\Theta^{ab}\Theta_{ab}+\Omega^{ab}\Omega_{ab}+\Theta^{2},\\
        8\pi{}P_{a}&=-2A^{b}\Omega_{ba}-D_{b}\tensor{B}{^{b}_{a}}+D_{a}\Theta,\\
        16\pi{}S&=-\tensor[^{\left(3\right)}]{R}{}-4\dot{\Theta}-3\Theta_{ab}\Theta^{ab}+3\Omega_{ab}\Omega^{ab}-\Theta^{2}+4D_{a}A^{a}+4A^{a}A_{a}\\
        8\pi{}S_{\left<ab\right>}&=\tensor[^{\left(3\right)}]{R}{_{\left<ab\right>}}+\dot\Sigma_{\left<ab\right>}-2\tensor{\Sigma}{^{c}_{\left<a\right.}}\Sigma_{\left.b\right>c}-2\tensor{\Omega}{^{c}_{\left(a\right.}}\Sigma_{\left.b\right)c}+\frac{1}{3}\Theta\Sigma_{ab}-D_{\left<a\right.}A_{\left.b\right>}-A_{\left<a\right.}A_{\left.b\right>}.
    \end{align}
    where $S$ is trace of $S_{ab}$.
    Meanwhile, modified Einstein equation $R_{ab}=8\pi\left(T_{ab}-\frac{1}{2}g_{ab}T\right)$ has splitting given by
    \begin{align}
        4\pi\left(E+S\right)&=-\dot{\Theta}-\Theta^{ab}\Theta_{ab}+\Omega^{ab}\Omega_{ab}+D_{a}A^{a}+A^{a}A_{a},\\
        4\pi\left(3E-S\right)&=\tensor[^{\left(3\right)}]{R}{}+\dot{\Theta}+\Theta^{2}-D_{a}A^{a}-A^{a}A_{a}
    \end{align}
    omitting redundant parts.

\section{Reducing to 3+1 Formalism}\label{sec:reducing_to_3+1+formalism}

3+1 formalism has different given structure which is a foliation of Cauchy hypersurfaces. In this case, the unit vector field $n$ normal to each hypersurface becomes a criterion for the splitting with the projection operator:
\begin{equation}
    \tensor{\left[\bot_{n}\right]}{^{a}_{b}}=\tensor{\delta}{^{a}_{b}}+n^{a}n_{b}.
\end{equation}
The most notable feature of this vector field is vanishing of vorticity by Frobenius theorem given as
\begin{equation}
    \Omega_{ab}=-\frac{1}{2}\bot_{n}\left(dn\right)_{ab}=\bot_{n}\nabla_{\left[b\right.}n_{\left.a\right]}=0.
\end{equation}
Hence, there is no temporal part in commutator between spatial vectors which reduces to
\begin{equation}
    \left[X,Y\right]^{a}=X^{b}D_{b}Y^{a}-Y^{b}D_{b}X^{a}.
\end{equation}
Then, the spatial connection becomes torsionless,
\begin{equation}
    \tensor[^{\left(3\right)}]{T}{^{a}_{bc}}=0,\\
\end{equation}
which implies
\begin{equation}
    D_{a}D_{b}f-D_{b}D_{a}f=0,
\end{equation}
from \Cref{eq:torsion_expansion}, where $f$ is an arbitrary function. Meanwhile, the two different definition of Riemann curvature \Cref{eq:naive_Riemann_curvature,eq:3d_riemann_curvature} become indentical and they reduce to
\begin{equation}
    \tensor[^{\left(3\right)}]{R}{^{a}_{bcd}}Z^{b}=D_{c}D_{d}Z^{a}-D_{d}D_{c}Z^{a}.
\end{equation}

The splitting of Riemann curvature associated with Levi-Civita Connection, i.e. Gauss, Codazzi, and Ricci relation, become
\begin{align}
    \bot_{u}\tensor{R}{^{a}_{bcd}}&=\tensor[^{\left(3\right)}]{R}{^{a}_{bcd}}+\tensor{\Theta}{^{a}_{c}}\Theta_{bd}-\tensor{\Theta}{^{a}_{d}}\Theta_{bc},\\
    \bot_{u}\left(\tensor{R}{^{a}_{bcd}}u^{b}\right)&=D_{c}\tensor{\Theta}{^{a}_{d}}-D_{d}\tensor{\Theta}{^{a}_{c}},\\
    \bot_{u}\left(\tensor{R}{^{a}_{bcd}}u^{b}u^{d}\right)&=D_{c}A^{a}+A^{a}A_{c}+\tensor{\Theta}{^{b}_{c}}\tensor{\Theta}{_{b}^{a}}-\tensor{\dot{\Theta}}{^{a}_{c}}.
\end{align}
From that, the first Bianchi identity implies
\begin{equation}
    \tensor[^{\left(3\right)}]{R}{_{a\left[bcd\right]}}=0.
\end{equation}
The derived Ricci tensor becomes symmetric as
\begin{equation}
    \tensor[^{\left(3\right)}]{R}{_{\left[ab\right]}}=0.
\end{equation}
Finally, the splitting of the Einstein equation becomes
\begin{align}
    16\pi{}E&=\tensor[^{\left(3\right)}]{R}{}-\Theta^{ab}\Theta_{ab}+\Theta^{2},\\
    8\pi{}P_{a}&=-D_{b}\tensor{\Theta}{^{b}_{a}}+D_{a}\Theta,\\
    16\pi{}S&=-\tensor[^{\left(3\right)}]{R}{}-4\dot{\Theta}-3\Theta_{ab}\Theta^{ab}-\Theta^{2}+4D_{a}A^{a}+4A^{a}A_{a}\\
    8\pi{}S_{\left<ab\right>}&=\tensor[^{\left(3\right)}]{R}{_{\left<ab\right>}}+\dot{\Sigma}_{\left<ab\right>}-2\tensor{\Sigma}{^{c}_{\left<a\right.}}\Sigma_{\left.b\right>c}+\frac{1}{3}\Theta\Sigma_{ab}-D_{\left<a\right.}A_{\left.b\right>}-A_{\left<a\right.}A_{\left.b\right>}.
\end{align}
The splitting of the modified Einstein equation might be more useful which is
\begin{align}
    4\pi\left(E+S\right)&=-\dot{\Theta}-\Theta^{ab}\Theta_{ab}+D_{a}A^{a}+A^{a}A_{a},\\
    4\pi\left(3E-S\right)&=\tensor[^{\left(3\right)}]{R}{}+\dot{\Theta}+\Theta^{2}-D_{a}A^{a}-A^{a}A_{a}.
\end{align}

\section{Discussion}

So far we have looked at the 1+3 formalism in a covariant way. In the formalism, the spacetime with a timelike congruence as a given structure is assumed. Tensors on the spacetime are split into temporal and spatial part according to the tangent direction to the congruence. The kinematical quantities defined by the natural derivatives are used to express the splitting of various derivatives. In particular, natural derivatives of vorticity are constrained by twice exterior derivatives of the unit tangent which vanishes.

The Lie derivative along the tangent and the spatial connection induced from the Levi-Civita connection are used to get temporal and spatial derivative, respectively. The splitting of covariant derivatives using them is the core process to develop the formalism. For the splitting of second-order covariant derivatives, there are total 8 parts, but we need only some part to develop the formalism.

Meanwhile, the spatial connection has non-vanishing torsion and the Riemann curvature tensor which is different to the usual definition. Consequently, the first Bianchi identity of the Riemann tensor and antisymmetric part of derived Ricci tensor does not vanish. One thing to keep in mind is that there are alternative definitions to the spatial Riemann curvature. It is an open problem how to define Riemann curvature tensor. Alternatives can be found at\cite{GEM}.

From the above results, the Einstein equation can be split into temporal and spatial parts. Moreover, if we choose the congruence whose direction is normal to a spacelike hypersurface, then the 1+3 formalism reduces to the 3+1 formalism. By doing so, we can have a deeper understanding of the 3+1 formalism. All these processes have performed in a covariant manner without the complexities caused by the introduction of a coordinate system or basis. Components of the results, if necessary, can be obtained easily because they are expressed in tensor.

Actually, this work was derived from a study which develops a covariant approach in spacetime perturbation with general background. In particular, in order to perform Lagrangian perturbation, a timelike congruence is needed as a given structure. (It usually corresponds to the trajectory of fluid.) Naturally, 1+3 splitting is appropriate for the problem. If the 1+3 splitting of Lagrangian perturbation is completed, it will have wide applicability such as astrophysics, cosmology, and post-Newtonian approximation. This will be the future work.

\begin{acknowledgments}
I would like to thank Sanghyeon Ahn, Yeong-Bok Bae, Gungwon Kang and Jinho Kim for helpful discussions and hospitalities. I was partially supported by the Basic Science Research Program through the National Research Foundation of Korea(NRF) funded by the Ministry of Education(NRF-2018R1D1A1B07041004).
\end{acknowledgments}

\bibliography{bib}

\begin{thebibliography}{10}
\expandafter\ifx\csname natexlab\endcsname\relax\def\natexlab#1{#1}\fi
\expandafter\ifx\csname bibnamefont\endcsname\relax
  \def\bibnamefont#1{#1}\fi
\expandafter\ifx\csname bibfnamefont\endcsname\relax
  \def\bibfnamefont#1{#1}\fi
\expandafter\ifx\csname citenamefont\endcsname\relax
  \def\citenamefont#1{#1}\fi
\expandafter\ifx\csname url\endcsname\relax
  \def\url#1{\texttt{#1}}\fi
\expandafter\ifx\csname urlprefix\endcsname\relax\def\urlprefix{URL }\fi
\providecommand{\bibinfo}[2]{#2}
\providecommand{\eprint}[2][]{\url{#2}}

\bibitem[{\citenamefont{Gourgoulhon}(2012)}]{Gourgoulhon2012}
\bibinfo{author}{\bibfnamefont{{\'E}.}~\bibnamefont{Gourgoulhon}},
  \emph{\bibinfo{title}{3+1 Formalism in General Relativity: Bases of Numerical
  Relativity}} (\bibinfo{publisher}{Springer Berlin Heidelberg},
  \bibinfo{address}{Berlin, Heidelberg}, \bibinfo{year}{2012}), ISBN
  \bibinfo{isbn}{978-3-642-24525-1},
  \urlprefix\url{https://doi.org/10.1007/978-3-642-24525-1\_1}.

\bibitem[{\citenamefont{Jantzen et~al.}()\citenamefont{Jantzen, Carini, and
  Bini}}]{GEM}
\bibinfo{author}{\bibfnamefont{R.~T.} \bibnamefont{Jantzen}},
  \bibinfo{author}{\bibfnamefont{P.}~\bibnamefont{Carini}}, \bibnamefont{and}
  \bibinfo{author}{\bibfnamefont{D.}~\bibnamefont{Bini}},
  \emph{\bibinfo{title}{Understanding spacetime splittings and their
  relationships or gravitoelectromagnetism: the user manual}},
  \urlprefix\url{http://www34.homepage.villanova.edu/robert.jantzen/gem/index.html}.

\bibitem[{\citenamefont{Roy}(2014)}]{Roy2014}
\bibinfo{author}{\bibfnamefont{X.}~\bibnamefont{Roy}}, \emph{\bibinfo{title}{On
  the 1+3 formalism in general relativity}} (\bibinfo{year}{2014}),
  \eprint{arXiv:1405.6319}.

\bibitem[{\citenamefont{Wald}(1984)}]{wald1984general}
\bibinfo{author}{\bibfnamefont{R.}~\bibnamefont{Wald}},
  \emph{\bibinfo{title}{General Relativity}} (\bibinfo{publisher}{University of
  Chicago Press}, \bibinfo{year}{1984}), ISBN \bibinfo{isbn}{9780226870335},
  \urlprefix\url{https://press.uchicago.edu/ucp/books/book/chicago/G/bo5952261.html}.

\bibitem[{\citenamefont{Jantzen et~al.}(1992)\citenamefont{Jantzen, Carini, and
  Bini}}]{JANTZEN19921}
\bibinfo{author}{\bibfnamefont{R.~T.} \bibnamefont{Jantzen}},
  \bibinfo{author}{\bibfnamefont{P.}~\bibnamefont{Carini}}, \bibnamefont{and}
  \bibinfo{author}{\bibfnamefont{D.}~\bibnamefont{Bini}},
  \bibinfo{journal}{Annals of Physics} \textbf{\bibinfo{volume}{215}},
  \bibinfo{pages}{1 } (\bibinfo{year}{1992}), ISSN \bibinfo{issn}{0003-4916},
  \urlprefix\url{http://www.sciencedirect.com/science/article/pii/000349169290297Y}.

\bibitem[{\citenamefont{Stewart}(1993)}]{stewart1993advanced}
\bibinfo{author}{\bibfnamefont{J.}~\bibnamefont{Stewart}},
  \emph{\bibinfo{title}{Advanced General Relativity}}, Cambridge Monographs on
  Mathematical Physics (\bibinfo{publisher}{Cambridge University Press},
  \bibinfo{year}{1993}), ISBN \bibinfo{isbn}{9780521449465},
  \urlprefix\url{http://www.cambridge.org/kr/academic/subjects/physics/cosmology-relativity-and-gravitation/advanced-general-relativity?format=PB\&isbn=9780521449465}.

\bibitem[{\citenamefont{Massa}(1974{\natexlab{a}})}]{Massa1974_1}
\bibinfo{author}{\bibfnamefont{E.}~\bibnamefont{Massa}},
  \bibinfo{journal}{General Relativity and Gravitation}
  \textbf{\bibinfo{volume}{5}}, \bibinfo{pages}{555}
  (\bibinfo{year}{1974}{\natexlab{a}}), ISSN \bibinfo{issn}{1572-9532},
  \urlprefix\url{https://doi.org/10.1007/BF02451398}.

\bibitem[{\citenamefont{Massa}(1974{\natexlab{b}})}]{Massa1974_2}
\bibinfo{author}{\bibfnamefont{E.}~\bibnamefont{Massa}},
  \bibinfo{journal}{General Relativity and Gravitation}
  \textbf{\bibinfo{volume}{5}}, \bibinfo{pages}{573}
  (\bibinfo{year}{1974}{\natexlab{b}}), ISSN \bibinfo{issn}{1572-9532},
  \urlprefix\url{https://doi.org/10.1007/BF02451399}.

\bibitem[{\citenamefont{Massa}(1974{\natexlab{c}})}]{Massa1974_3}
\bibinfo{author}{\bibfnamefont{E.}~\bibnamefont{Massa}},
  \bibinfo{journal}{General Relativity and Gravitation}
  \textbf{\bibinfo{volume}{5}}, \bibinfo{pages}{715}
  (\bibinfo{year}{1974}{\natexlab{c}}), ISSN \bibinfo{issn}{1572-9532},
  \urlprefix\url{https://doi.org/10.1007/BF00761928}.

\bibitem[{\citenamefont{Boersma and Dray}(1995)}]{boersma1995}
\bibinfo{author}{\bibfnamefont{S.}~\bibnamefont{Boersma}} \bibnamefont{and}
  \bibinfo{author}{\bibfnamefont{T.}~\bibnamefont{Dray}},
  \bibinfo{journal}{Journal of Mathematical Physics}
  \textbf{\bibinfo{volume}{36}}, \bibinfo{pages}{1378} (\bibinfo{year}{1995}),
  \eprint{https://doi.org/10.1063/1.531127},
  \urlprefix\url{https://doi.org/10.1063/1.531127}.

\end{thebibliography}

\end{document}